# First observation of antimatter wave interference

A. Ariga[1], A. Ereditato[1], R. Ferragut[2,3], M. Giammarchi[3]*, M. Leone[2], C. Pistillo[1], S. Sala[4,3], P. Scampoli[1,5]

**In 1924 Louis de Broglie introduced the concept of wave-particle duality[1]: the Planck constant $h$ relates the momentum $p$ of a massive particle to its de Broglie wavelength $\lambda_{dB} = h/p$. The superposition principle is one of the main postulates of quantum mechanics; diffraction and interference phenomena are therefore predicted and have been observed on objects of increasing complexity, from electrons[2,3] to neutrons[4] and molecules[5]. Beyond the early electron diffraction experiments[2,3], the demonstration of single-electron double-slit-like interference was a highly sought-after result. Initially proposed by Richard Feynman as a thought experiment it was finally carried out in 1976[6]. A few years later, positron diffraction was first observed[7]. However, an analog of the double-slit experiment has not been performed to date on any system containing antimatter. Here we present the first observation of matter wave interference of single positrons, by using a period-magnifying Talbot-Lau interferometer[8] based on material diffraction gratings. Individual positrons in the 8-14 keV energy range from a monochromatic beam were detected by high-resolution nuclear emulsions[9]. The observed energy dependence of fringe contrast proves the quantum-mechanical origin of the detected periodic pattern and excludes classical projective effects. Talbot-Lau interferometers are well-suited to the experimental challenges posed by low intensity antimatter beams and represent a promising option for measuring the gravitational acceleration of neutral antimatter.** The experiment makes use of the variable energy positron beam facility of the L-NESS laboratory in Como (Italy). Positrons (e$^+$) from the beta decay of a $^{22}$Na radioactive source are implanted on a monocrystalline tungsten film and emitted with a kinetic energy of about 3 eV, determined by the work function of the material[10]. Slow positrons are then accelerated up to 16 keV by means of a purely electrostatic system. A monochromatic and continuous beam is thus formed with energy spread less than 0.1 %, limited by the stability of the power supplies. The positron flux is $(5 \pm 1) \times 10^3$ e$^+$/s and beam focusing can be tuned to reach spot sizes of the order of several millimeters of FWHM, an approximately Gaussian intensity profile[11] and an angular divergence at the level of a few millirad[9]. A suitable model[12] of grating-based interferometry with a partially coherent beam exploits the analogy with Gaussian Schell-model beams of classical optics. Within this formalism incoherence translates physically into a broad transverse momentum distribution and mathematically to a short transverse coherence length $l$. The L-NESS beam features a coherence length of the order of a few nanometers. An effective configuration for these conditions was obtained with a novel period-magnifying two-grating interferometer[8]. It exploits an intermediate working regime between the standard Talbot-Lau setup, where the two gratings and the detector are equally spaced, and the so-called Lau interferometer[13], which has more stringent coherence requirements[12]. By means of unequal grating periodicities the system provides sizable period magnification in a relatively compact setup[8]. In particular, we employed gold coated 700 nm thick silicon nitride (SiN) gratings with periodicity $d_1 = (1.210 \pm 0.001)$ μm and $d_2 = (1.004 \pm 0.001)$ μm to produce a $d_3 = (5.90 \pm 0.04)$ μm periodic interference pattern. Both gratings have a nominal open fraction of 50%.

[1]Albert Einstein Center for Fundamental Physics, Laboratory for High Energy Physics, University of Bern, Bern, Switzerland. [2]L-NESS Laboratory, Politecnico di Milano, Como, Italy. [3]Istituto Nazionale di Fisica Nucleare (INFN), Sezione di Milano, Milano, Italy. [4]Dipartimento di Fisica "Aldo Pontremoli", Università degli Studi di Milano, Milano, Italy. [5]Dipartimento di Fisica "Ettore Pancini", Università di Napoli Federico II, Complesso Universitario di Monte S. Angelo, Napoli, Italy. *e-mail: marco.giammarchi@mi.infn.it

The periodic spatial distribution generated by the interferometer (Fig. 1) is revealed by a nuclear emulsion detector. Nuclear emulsions[14] offer submicron level position resolution in the detection of ionizing particles[9,15]. They work as photographic films by exploiting the properties of silver-bromide crystals embedded in a 50 μm thick gelatin matrix. For this experiment we developed a glass-supported emulsion detector and experimentally demonstrated its capability to resolve periodic patterns at the micrometric scale even with low signal contrast and on large areas[9].

Interferometer alignment is particularly challenging and it is intimately connected to beam coherence. The Talbot-Lau interferometer can produce high-contrast fringes if the resonance condition[8]

$$\frac{L_1}{L_2} = \frac{d_1}{d_2} - 1$$

is met, even for a fully incoherent ($l \to 0$) beam. However, as this regime is approached the required accuracy at which the above condition should be ensured increases. In order to improve the tolerance to possible misalignments the beam was collimated by means of two circular openings (Fig. 1). Regardless of beam coherence the experimental uncertainty on the optimal (resonance) value of the ratio $\frac{L_1}{L_2}$ that stems from the errors on the measured grating periods amounts to $\sigma_{L_1/L_2} = 0.002$. In the adopted geometry (Fig. 1) this corresponds to an uncertainty on the ideal detector location $L_2$ at the level of 5 mm. To circumvent this issue we operated emulsion films tilted by 45° in such a way that the $Y$ coordinate of the emulsion plane is correlated to $L_2$, which varies along the z-axis in the laboratory reference frame (Fig. 1). A position scan was thus performed in a single exposure by analyzing different horizontal slices of the emulsion detector.

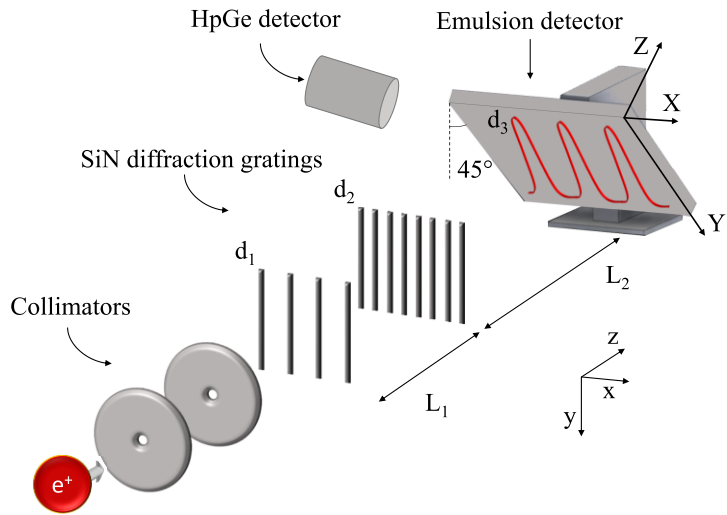

Fig. 1 | **Schematics of the Talbot–Lau interferometer**. Positrons traverse two circular 2 mm wide collimators 10.2 cm apart. The interferometer is composed of two SiN diffraction gratings with periodicity $d_1$ and $d_2$, respectively, separated by $L_1 = (118.1 \pm 0.2)$ mm. Interference fringes with $d_3$ periodicity are expected at $L_2 = (576 \pm 5)$. The emulsion is tilted so that the $Y$-axis in the reference frame of the emulsion surface $(X,Y)$ forms a 45° angle with the y-axis of the laboratory. Gamma-rays from positron annihilation (511 keV) in the emulsion are monitored with a high-purity germanium (HpGe) detector for flux measurement.

Relative rotational alignment of the two gratings is a critical parameter, as contrast follows a Gaussian modulation as a function of the ϕ angle formed by the slits, with standard deviation[12]

$$\sigma_\phi = \frac{1}{\sqrt{2\pi}} \frac{d_2 l^{det}}{L_2 \lambda_{dB}}$$

where $l^{det}$ is the coherence length computed at the detector plane by an analytical model[12]. Throughout the measurements the ratio $l^{det}/\lambda_{dB} \sim 800$ remained approximately constant, which yielded a tolerance

$\sigma_\phi \sim$ 550 μrad for the collimated beam. The second grating was mounted on a piezoelectric rotation stage with nominal resolution of 0.8 μrad; the adopted alignment protocol ensures that $\phi <$ 70 μrad for the full duration of the exposures. All components of the interferometer, including the rotation stage, were non-magnetic and the interferometer was surrounded by a mu-metal shield designed to reduce the residual Earth magnetic field to less than 0.5 μT. Parallelism between the interferometer optical axis and the beam propagation axis was controlled at the level of 1 mrad by means of alignment lasers. The apparatus operated at a vacuum pressure between $10^{-7}$ and $10^{-6}$ mbar during each measurement.

The interferometer was configured for maximum contrast at E = 14 keV ($\lambda_{dB}$ = 10.3 pm) by setting[8] $L_1 = \frac{d_1 d_2}{\lambda_{dB}}$. Since this geometry satisfies Talbot-Lau resonance conditions, single-slit diffraction from the first to second grating plane is not negligible (this occurs because $\frac{L_1 \lambda_{dB}}{d_1} > d_2$). Therefore, a quantum-mechanical description of the system is required, which predicts a peculiar contrast dependence on the positron energy. To ensure uniform working conditions throughout the explored energy range we employed nuclear emulsions prepared without the standard surface protective gelatin layer. The presence of even a micrometric layer would have introduced a detection efficiency[11] and a positron trajectory smearing[9] that are both energy dependent. Thermally-induced grains are the dominant source of background noise in emulsion detectors. Nonetheless, no measurable increase in the average density of noise grains was observed for the unprotected emulsions compared to standard detectors used in the same conditions.

We now summarize the results of the four beam exposures performed for positron energies of 14, 11, 9 and 8 keV. This range was chosen to emphasize the maximum contrast drop while keeping a still detectable minimum contrast. Since the transport efficiency through the L-NESS electrostatic system decreases rapidly below 8 keV, lower energies would have required unpractically long exposure times. On the other hand, energies higher than 15 keV would have run into power supply stability and vacuum discharge limitations. The residual beam flux at the detector after collimation (~80% loss) and passage through the gratings (~90% loss) was at the level of 100 positrons per second. In order to accumulate sufficient statistics exposure times between 120 and 200 hours were required. These were selected to match the total counts on the HpGe detector measured for 14 keV. Beam focusing was tuned to ensure that beam spot size and hence the geometrical features of the beam such as angular divergence, did not appreciably vary with the energy. Spot sizes deviated by less than 10% from the average value of 6.5 mm FWHM on the detector plane. Silver-bromide crystals activated by the passage of the positrons through the emulsion become visible with optical microscopes after chemical development; emulsions were then processed at the microscope scanning facility of the University of Bern. For each given *X-Y* position the microscope grabbed a series of images by shifting the focal plane in the direction normal to the emulsion surface (*Z*). Clusters were reconstructed[16] to assign *X,Y,Z* coordinates to the positron impact point. The data were subdivided in 370 × 294 μm² wide regions, which we refer to as *views*. Each view was independently analyzed to search for periodicity in the distribution of points by maximizing the so-called *Rayleigh test* function[17]

$$R(\alpha, d_3) = \left| \frac{1}{n} \sum_{j=1}^{n} exp\left( \frac{2 \pi i X_j(\alpha)}{d_3} \right) \right|$$

with respect to the period $d_3$ and the angle $\alpha$ between the fringes and the microscope reference frame. We introduced $X_j(\alpha)$ as the *X*-coordinate of the *j*-th grain in the view rotated by $\alpha$. The effectiveness of this period-finding approach was experimentally demonstrated for similar applications[9,18]. Values of the parameters maximizing R, namely $(\alpha^*, d^*_3)$, were searched in the range $-0.05$ rad $< \alpha < 0.05$ rad and

5.7μm < $d_3$ < 6.1μm. If a consistent periodic signal spanned a large area, a distinctive peak would be expected in the $(\alpha^*, d_3^*)$ distribution over the analyzed views[9]. A representative example is shown in Fig. 2 for the 14 keV exposure.

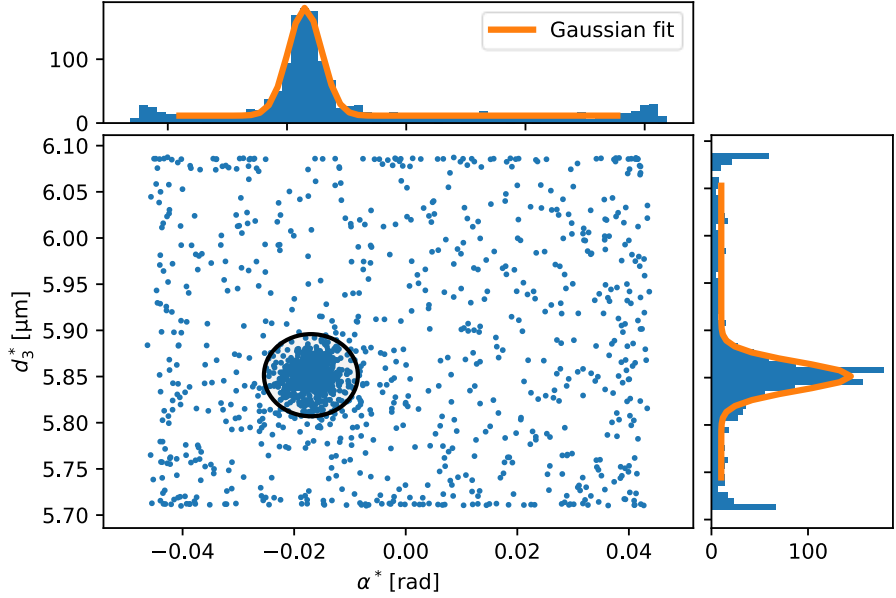

Fig. 2 | **Optimal angle and period found via the Rayleigh test.** Scatter plot and profile histograms (60 bins) of the optimal period and angle $(\alpha^*, d_3^*)$ for E=14 keV. A total of 1620 views covering a surface of about 10 × 14 mm² were analyzed. Histograms are fit with a Gaussian function plus a constant background. The black ellipse indicates the corresponding 99.7% C.L. region. The same scanning surface was considered for all energies.

A least squares Gaussian fit to the data (Fig. 2) yields the detected period for each exposure:

$$d_3^{(14)} = (5.851 \pm 0.001_{stat} \pm 0.050_{syst}) \text{ μm}$$

$$d_3^{(11)} = (5.852 \pm 0.001_{stat} \pm 0.050_{syst}) \text{ μm}$$

$$d_3^{(9)} = (5.854 \pm 0.001_{stat} \pm 0.050_{syst}) \text{ μm}$$

$$d_3^{(8)} = (5.850 \pm 0.001_{stat} \pm 0.050_{syst}) \text{ μm}$$

The statistical error (68% C.I.) comes from the fit procedure, whereas the estimated 0.8% systematic error descends from the conversion between camera pixels and physical size[9]. The measured period is compatible with the expected value $d_3 = (5.90 \pm 0.04)$ μm. Once the optimal period and angle were found, a histogram of $X_j(\alpha^*)$ mod $d_3^*$ was constructed, where mod is the *modulo operation* on floating point numbers. The signal contrast was estimated by fitting the histogram with a $d_3^*$-periodic sinusoidal function appropriate for the phase structure of the fringes (Fig.3a). Subtraction of a constant background from the histogram was performed by measuring the density of grains in the bulk, as positrons can only penetrate a few microns of emulsion. This yielded an estimate of the number of grains due to intrinsic emulsion noise entering the analysis region. This is limited to 340 × 270 μm² on the X-Y plane to remove areas affected by sizeable optical aberration. The depth in Z was determined with a Gaussian fit on the positron implantation profile[9]. The average width of the selected region along the Z direction was 2.9 μm. This procedure was performed only for views contained in the 3σ elliptical region defined in Fig. 2. As the signal-to-noise ratio decreased moving away from the center of the beam spot, maximization of the Rayleigh test essentially converged to random values $(\alpha^*, d_3^*)$ within the search region; contrast estimation for these views would give unphysical results.

The contrast measured after noise subtraction is in principle independent from the signal-to-noise ratio. An average noise density of 5.8, 7.0, 7.4 and 6.2 grains/1000 μm$^3$ was measured for the three energies, respectively, with a standard deviation of 0.5 grains/1000 μm$^3$ in all cases. The average number of grains entering the analysis was within 10% of 11000 for all exposures. A two-dimensional heatmap of the measured contrast covering the scanned surface area is shown for the four energies in Fig. 3b. The contrast of views excluded from the analysis was set to zero. Since $Y$ is correlated to $L_2$, a contrast modulation was observed in the $Y$ direction. The contrast dependence on $X$ displayed a marked asymmetry, probably as a result of limited beam coherence and alignment accuracy. We emphasize that this feature would have been hidden in a moving-mask based detection scheme. On the other hand, emulsions allowed direct fringe detection on an area significantly larger than the surface of the gratings (3 × 3 mm$^2$): the scanned area contained a sizable number (~10$^4$) of consecutive high visibility $d_3$-periodic fringes. Visual inspection of the two-dimensional maps already suggests that contrast is decreasing with the energy. In order to provide a quantitative estimate of the peak contrast we selected views in a 1 mm wide region (indicated by the dashed lines in Fig. 3b). The corresponding contrast is shown in Fig. 3a as a function of the position of the geometrical center of the view.

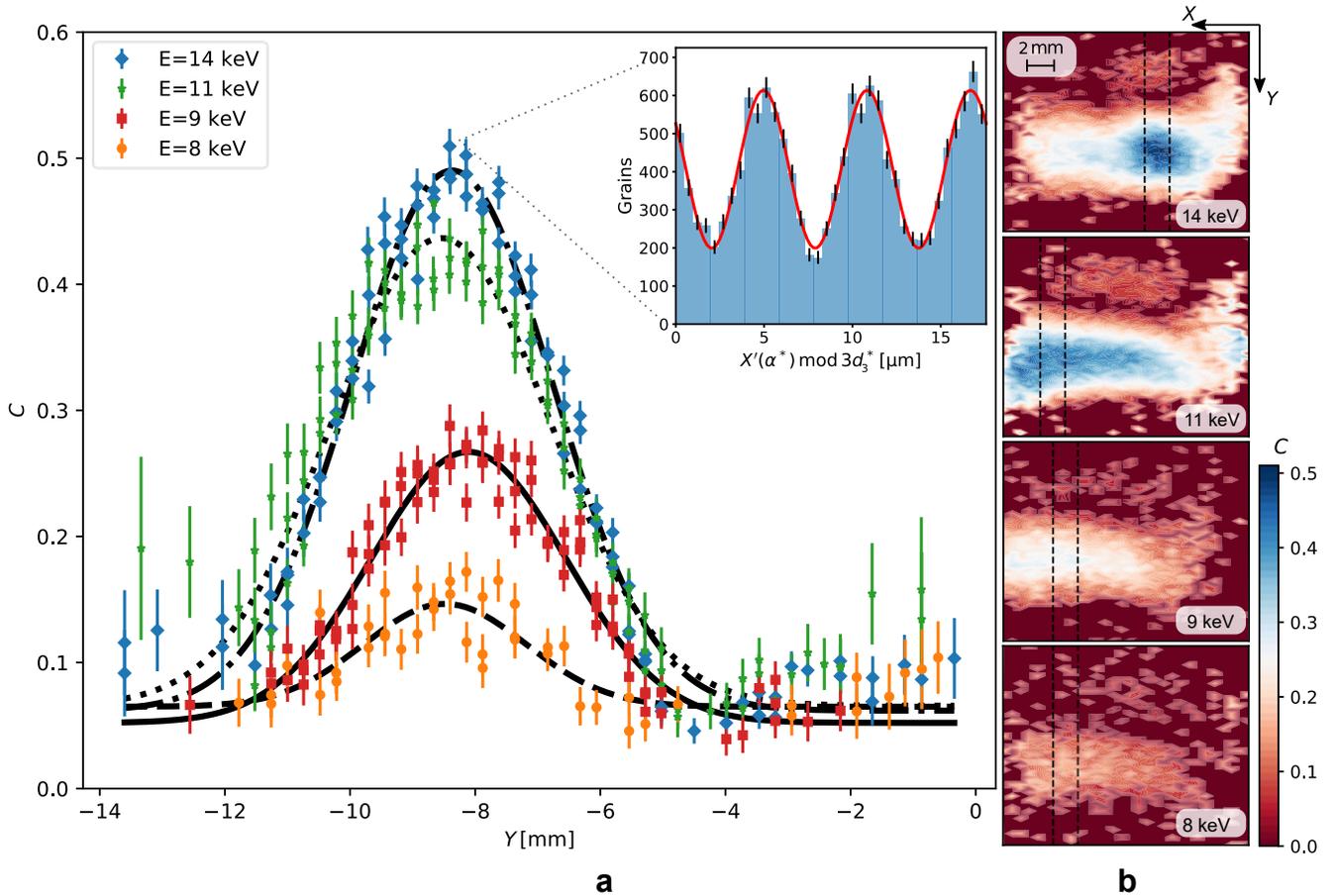

Fig. 3 | **Summary of the results. a**, contrast $C$ as a function of $Y$ for views in the region delimited by the dashed lines in **b**. Errors come from a sinusoidal least-squares fit. The result of a Gaussian fit with constant background is superimposed to the data. The inset shows a histogram of $X_j(\alpha^*) \bmod 3d^*$ and a sinusoidal fit for the highest contrast view; error bars represent Poissonian counting uncertainties. **b**, contrast heatmaps for the four energies considered.

The data were fit with a Gaussian function plus a constant background (Fig. 3a), which was used to estimate the maximum contrast for each energy:

$$C^{(14)} = 0.491 \pm 0.005$$
$$C^{(11)} = 0.436 \pm 0.008$$
$$C^{(9)} = 0.267 \pm 0.008$$
$$C^{(8)} = 0.144 \pm 0.008$$

where 68% C.I. uncertainties were estimated from the fit parameters. The position of the maximum was also determined from the fit:

$$Y_0^{(14)} = -8.4 \pm 0.5 \text{ mm}$$
$$Y_0^{(11)} = -8.5 \pm 0.5 \text{ mm}$$
$$Y_0^{(9)} = -8.1 \pm 0.5 \text{ mm}$$
$$Y_0^{(8)} = -8.5 \pm 0.5 \text{ mm}$$

where uncertainties come from experimental errors in positioning the emulsion film onto the reference frame of the microscope and in the interferometer.

We conclude that the peak contrast of the periodic signal was observed for different energies at the same distance $L_2 = (573 \pm 1)$ mm, where the uncertainty accounts for experimental errors in the position of the emulsion in the laboratory reference frame. This measured contrast modulation is in agreement with the expected behavior for a Talbot-Lau interferometer and is incompatible with classical moiré deflectometry[19], where particles propagate on ballistic trajectories through the gratings. Since this geometry could in principle produce a fringe pattern with the same periodicity from geometrical shadow effects alone[8], this conclusion holds only if spurious energy-dependent effects are ruled out. A possible source could be the energy-dependent positron transmission from the grating bars, which varies from about 0.1% at 8 keV to 49% at 14 keV. This contribution to background noise is however suppressed by a factor $10^4$ with respect to the intrinsic emulsion noise due to the broad angular distribution of transmitted positrons. The measured contrast drop from the resonance value, defined as $C/C_{max}(E)$, is shown in Fig. 4 as a function of the energy and compared with the quantum-mechanical and classical predictions. As the resonance plane was identified by means of the tilted detector, we assumed ideal longitudinal alignment in the analytical calculation. We emphasize that the model contains several simplifying assumptions and is thus only suitable for a qualitative comparison with the data.

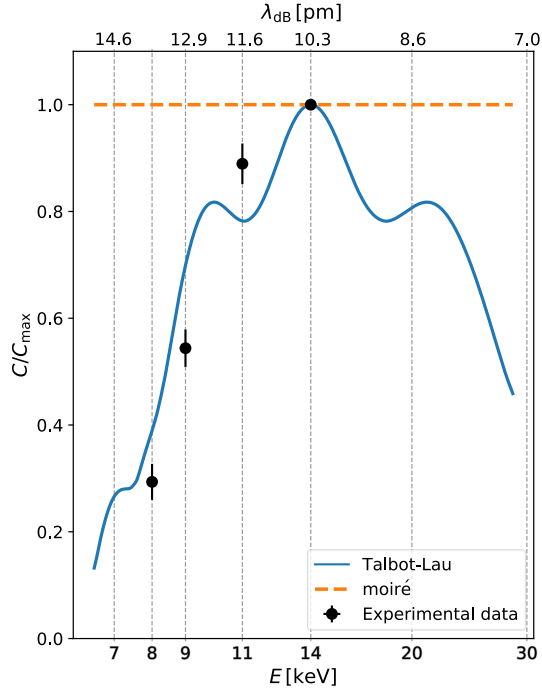

Fig. 4 | **Contrast as a function of energy.** Measured contrast normalized to the resonance value, defined as C/C$_{max}$(E). The 68% C.I. uncertainties are obtained by standard error propagation. The solid line is the quantum-mechanical prediction, while the classical prediction is indicated by the dashed line.

The evidence described in this paper allows to conclude that we successfully observed positron interference in the Talbot-Lau regime. Considering the six orders of magnitude difference between the typical transit time through the interferometer (10 ns) and the average time distance between two consecutive positrons (10 ms), a genuine single-particle experiment was realized. Echoing the Merli[6] and Tonomura[20] "double-slit" electron experiments, our measurement illustrates the principle of wave-particle duality: positrons are emitted as point-like particles by a radioactive source, they interact as de Broglie waves with the interferometer, and are eventually detected as distinct spots in the emulsion detector. Their spatial distribution is driven by $\lambda_{dB}$ as predicted by quantum mechanics, providing the first demonstration of antimatter wave interferometry. This result is the first step in the QUPLAS (QUantum interferometry with Positrons and LASers) program[8,9,11,21] and paves the way to interferometric studies of fundamental interest on systems containing antimatter. In particular, the related topic of antimatter gravity[22–29] can profit from the capability to work with low intensity, incoherent beams. Therefore, Talbot-Lau based inertial sensing[8] is a promising technique to tackle the measurement of the gravitational acceleration antimatter systems such as positronium, antihydrogen or muonium.

**Acknowledgements.** The authors would like to thank S. Cialdi and M. Potenza for their contribution to the laser alignment systems. We acknowledge the support received from S. Aghion, G. Maero and M. Romè on beam operations. We are grateful to S. Olivares and F. Castelli for their theoretical insights. We are in debt to T. Ariga for sharing her expertise in emulsion production. M. Bollani and M. Lodari are acknowledged for useful discussions. We thank R. Haenni for his efforts with machining work and M. Vladymyrov who developed the data acquisition software. We thank T. Savas for the fabrication of the diffraction gratings. We also thank L. Miramonti, P. Lombardi and G. Ranucci for providing useful equipment. Financial support from the Politecnico di Milano, the INFN and the Laboratory for High Energy Physics of the University of Bern is also acknowledged.